%% file: main.tex
\documentclass[a4paper]{article}

\usepackage{INTERSPEECH2020}
\usepackage{booktabs}
\usepackage{threeparttable}

\title{Deep Learning Based Assessment of Synthetic Speech Naturalness}
\name{Gabriel Mittag$^1$, Sebastian M\"oller$^{1,2}$}
\address{$^1$Quality and Usability Lab, Technische Universit\"at Berlin, Berlin, Germany \\
$^2$Deutsches Forschungszentrum f\"ur K\"unstliche Intelligenz (DFKI), Berlin, Germany }
\email{gabriel.mittag@tu-berlin.de, sebastian.moeller@tu-berlin.de}

\begin{document}

\maketitle
\begin{abstract}
In this paper, we present a new objective prediction model for synthetic speech naturalness. It can be used to evaluate Text-To-Speech or Voice Conversion systems and works language independently. The model is trained end-to-end and based on a CNN-LSTM network that previously showed to give good results for speech quality estimation. We trained and tested the model on 16 different datasets, such as from the Blizzard Challenge and the Voice Conversion Challenge. Further, we show that the reliability of deep learning-based naturalness prediction can be improved by transfer learning from speech quality prediction models that are trained on objective POLQA scores. The proposed model is made publicly available and can, for example, be used to evaluate different TTS system configurations.
\end{abstract}
\section{Introduction}
In the evaluation of Text-To-Speech (TTS) systems, one of the main performance indicators is how natural the synthesized speech sounds. To measure the naturalness, auditory listening tests are performed. In these tests, participants rate the naturalness on a five-point absolute category scale. The average across all test participants then gives the naturalness MOS (mean opinion score). Although this evaluation is time-consuming and costly, so far, there is no reliable instrumental model available. In contrast, for the assessment of communication networks full-reference speech quality models, such as PESQ and POLQA \cite{P863}, have been established as objective MOS measures. They compare the clean reference with the degraded output speech signal and use distance features between both signals to estimate a MOS value. However, they cannot be applied to predict synthesized speech naturalness as these models focus on distortions caused by the transmission channels and are not trained on speech naturalness. Also, in the case of TTS evaluation, often, there is no reference speech available. To overcome this problem in the speech quality domain, non-intrusive models that only rely on the degraded output signal, such as P.563 \cite{P563} or NISQA \cite{mittag_a}, have been developed, but again, they also cannot be applied to synthesized speech directly.

A great resource for synthesized speech samples with naturalness ratings from listening tests are the annually held Blizzard Challenges \cite{blz_decade}. In this since 2005 held challenge, different teams build a TTS system from the same given source speech material. Each team then synthesizes a prescribed set of test sentences, which are evaluated in auditory listening tests. In a TTS naturalness dataset there are usually several sentences of the same system available. It is, therefore, a common approach to evaluate synthetic speech naturalness prediction models on a per-stimuli (i.e. per-file) level and additionally also on a per-system level. The per-system level evaluation is conducted by calculating the average prediction results per system and compare it with the average listening ratings per system.

In the following, we will give a brief overview of previous work on the instrumental assessment of synthesised speech naturalness or TTS quality. In \cite{falk_tts} a model was presented that took upon ideas in \cite{mariniak1993global} to compare perception-based features of synthesized speech with features extracted from natural speech. In \cite{falk2008improving} it was shown that the single-ended model P.563 is not suitable for directly predicting the quality of TTS systems. However, more promising results could be obtained when combining internal P.563 features with perception-based distance features \cite{moller2010comparison}, where a per-system Pearson's correlation coefficient (PCC) of 0.70/0.77 \cite{hinterleitner2010comparison} could be achieved on the Blizzard Challenge data of 2018/2019. In \cite{hinterleitner2013predicting} a TTS quality model that uses a large-scale of different features is proposed and in \cite{NORRENBROCK201517} a model based on prosodic and MFCC features that predicts perceptual quality dimensions of TTS systems achieved promising results. In \cite{latacz2015double} a double-ended speech naturalness model was presented that uses the English Blizzard Challenge data from 2008-2013 for training and evaluating with a per-system PCC of 0.84. In \cite{soni2016non} a naturalness prediction model based on spectral features has been proposed that obtained PCCs between 0.69 and 0.89 on the Blizzard Challenge data from 2008-2010 and 2012. More recently, also, neural networks have been used to predict the naturalness of synthesized speech. In \cite{yoshimura2016hierarchical} different models (linear regression, feed-forward and convolutional neural networks) together with MFCC and P.563 features as input, have been used to predict the naturalness again on the English Blizzard Challenge data from 2008-2013 (PCC of 0.74). In \cite{tang2019text} a LSTM (long short-term memory) network with MFCC and P.563 features was presented that was trained and evaluated on the Mandarin data from the Blizzard Challenges 2008, 2009, and 2010, on which it obtained a PCC of 0.68. 

The first deep learning-based approach was presented in \cite{automos}, where a model based on LSTM networks with mel-spectrograms as inputs is proposed that achieved a per-system PCC of 0.93 on internal datasets. In \cite{guo2020naturalness} a model that uses residual CNN networks is proposed and showed to outperform LSTM networks on the Mandarin Blizzard challenge data from 2008-2010. MOSnet \cite{Lo2019} is a CNN-LSTM based voice conversion objective assessment model that is similar to the proposed model in this paper. It is trained and validated on the Voice Conversion Challenge data of 2018, where it obtained a PCC of 0.96 on the validation set and a PCC of 0.92 on the VCC 2016 data, which was used as a test set.

In this paper, we present a new synthetic speech naturalness prediction model based on a CNN-LSTM network architecture with transfer learning domain knowledge from a speech quality database. We collected 16 datasets with synthesized speech and naturalness ratings that contain 12 different languages for training and evaluating. The proposed model is language independent and works across TTS synthesizers and voice conversion models. Further, it should be noted that in most of the previous work on instrumental TTS evaluation, cross-validation approaches are used to test the models, where speech samples from the same dataset are used for training and testing. To see how our model generalizes on unseen data, we validate and test it on datasets that were not used for training. The proposed model is made publicly available to evaluate TTS systems and to improve the naturalness prediction further \footnote{github.com/gabrielmittag/NISQA}.

\section{Model}
The model is the same CNN-LSTM network that has been used for single-ended speech quality estimation in \cite{mittag_b}. At first mel-spectrograms are calculated from the speech waveform. To make the application of the model for speech files with different sample rates easier, we use a fixed FFT size of 4048 samples with an adaptive window length of 20\,ms and hop length of 10\,ms. In this way, down or upsampling of speech files to a certain sample rate is not needed. We decided to use 48 mel bands with a maximum frequency of 8\,kHz because most files in the training samples have a sample rate of 16\,kHz. Also, we did not normalize the speech levels of the input signals on purpose, so that the model would be able to learn to handle different speech levels automatically and to avoid another preprocessing step when the model is applied for prediction.

The mel-spectrograms are then divided into spectrogram segments with a duration of 150\,ms and a hop length of 10\,ms that are used as input to the CNN network, yielding an input size of 48x15. The output of the CNN is a feature vector of size 20 for each spectrogram segment. The sequence of CNN feature vectors is then used as input for a "many-to-one" bidirectional LSTM network with one layer and 128 hidden units that models time dependencies and estimates the overall speech naturalness.

The CNN network consists of overall 6 convolutional layer with 16, 32, and 4x64 filters. The model design is outlined in Table \ref{tab:cnn}, where $N$ is the sequence length that depends on the signal duration. For more details of the neural network also see the open-sourced code.
\begin{table}[htb]
\caption{CNN design (each convolutional layer is followed by a batch normalization and ReLu layer. The kernel size is 3x3.)}
\vspace{-0.3cm}
\centering
\label{tab:cnn}
\resizebox{0.65\linewidth}{!}{%
\begin{tabular}{l|l}
\textbf{Layer}         & \textbf{Output size} \\ \hline
Input                  & $N$x1x48x15  \\ \hline
Conv 1                 & $N$x16x48x15 \\
Pool                   & $N$x16x24x8 \\ \hline
Conv 2                 & $N$x32x24x8 \\
Pool / Dropout(20\%)   & $N$x32x12x4   \\ \hline
Conv 3                 & $N$x64x12x4   \\
Conv 4                 & $N$x64x12x4   \\
Pool / Dropout(20\%)   & $N$x64x6x2  \\ \hline
Conv 5 / Dropout(20\%) & $N$x64x6x2   \\
Conv 6                 & $N$x64x6x2   \\ \hline
FC                     & $N$x20      
\end{tabular}
}
\end{table}
\section{Datasets}
\subsection{Speech quality pretraining}
We use the pretraining database from \cite{mittag_b} to first build a speech quality prediction network that is trained on speech communication network degradation. We then use the speech quality prediction domain knowledge of the neural network to improve the reliability of synthesized speech naturalness prediction through transfer learning. The database contains 100,000 speech files that are based on 5,000 English and German reference speech files from the AusTalk\footnote{https://bigasc.edu.au/} \cite{estival2014austalk} and NSC \cite{NSC} corpus. Overall the reference sentences come from over 1,000 different speakers. We then simulated a wide variety of different distortions, such as codecs (G711, G722, AMR-NB, AMR-WB, Opus, EVS) with different bitrates, sbackground noises, white noise, amplitude clipping, time clipping, packet-loss, frequency filter, and combinations of these distortions. We, in particular, suspect the codec and packet-loss conditions to be helpful for speech naturalness predictions since packet-loss concealment algorithms synthesize new speech frames when a packet is lost during transmission. This synthesis often leads to artificial sounds that can be similar to TTS synthesized speech. Because this large database has no subjective ratings available, we predict MOS values with POLQA that we use for model training.
\subsection{Blizzard}
We used all available results of the Blizzard Challenge\footnote{www.cstr.ed.ac.uk/projects/blizzard/data.html} \cite{blz08, blz09, blz10, blz11, blz12, blz13, blz14, blz15, blz16, blz19}, which are currently all years except for 2017 and 2018 (see Table \ref{tab:blz}). For three years (2014, 2015, 2016) no raw rating results but only per-system MOS are available. Therefore, we used all available stimuli and assigned them with their system MOS. For all other years, the individual ratings per file are used for training and evaluation. On average,  there are 13 ratings per file available. Each year had 8-24 challenge entries. Additionally, natural speech and a baseline model were part of the auditory listening tests. Most of the challenges had different tasks, such as a main hub task with all available training data, another hub task with fewer training samples, and different spoke sub tasks. We divided each year by task and speaker and treated them as separate datasets for the model evaluation.
\begin{table}[htb]
\caption{Blizzard Challenges overview}
\vspace{-0.3cm}
\centering
\label{tab:blz}
\resizebox{0.98\linewidth}{!}{%
\begin{tabular}{c|c|c|c|c}
Year & Speaker          & Lang                   & \# Files & \# Teams \\ \hline
2008 & Roger, CAS       & en, zh                 & 1942     & 19         \\
2009 & Roger, iFLYTEK   & en, zh                 & 2056     & 19         \\
2010 & RJS, Roger, CAS  & en, zh                 & 2638     & 17         \\
2011 & Nancy            & en                     & 507      & 9          \\
2012 & John Greenman    & en                     & 442      & 9          \\
2013 & Catherine Byers  & en                     & 1356     & 14         \\
2014 & 6 Indian speaker & as, gu, hi, rj, ta, te & 8990     & 9          \\
2015 & 6 Indian speaker & bn, ml, hi, mr, ta, te & 5200     & 8          \\
2016 & Lesley Sims      & en                     & 1156     & 13         \\
2019 & Zhenyu Luo       & zh                     & 1352     & 24        
\end{tabular}
}
\end{table}
\subsection{Voice Conversion Challenge}
The Voice Conversion Challenge \cite{vcc1, vcc2, vcc3} was held the first time in 2016 and a second time in 2018. The task of this challenge is to transform a speaker identity included in a source speech waveform into a different one while preserving linguistic information of the source speech. 

The material of the 2018 challenge is publicly available\footnote{https://datashare.is.ed.ac.uk/handle/10283/3061} and consisted of two different tasks. In the hub task a voice conversion model is built from a parallel clean training database, where the source and target speaker read out the same set of utterances. In the spoke task a non-parallel database was used for training, in which source and target speaker read out different sentences. While voice-conversed speech is not produced by TTS systems, they lead to similar speech degradation and are also evaluated in terms of speech naturalness. Overall, 20,874 files are available in this dataset, which is much larger than the other TTS datasets used in this paper. To avoid an imbalanced model training on the VCC files, we randomly subsampled 1000 files each for the hub and spoke challenge for training and another 1000 files each for model validation. Each file was on average rated by four listening test participants. The speech samples used in the listening test consist of 16 speaker-pairs (four source speakers and four target speakers) from 23 different participating teams. 

The material from the first challenge in 2016 is also available\footnote{https://datashare.is.ed.ac.uk/handle/10283/2211}; however, it does not contain per-stimuli ratings. While the speech samples came from the same dataset as VCC 2018, the speakers of the test speech samples were different. Because no per-stimuli ratings are available, we use this year as a test set. Also, this year has been used as a test for the MosNet evaluation as well, which makes it easier to compare the performance of both models.
\input{results_table.tex}
\subsection{TU Berlin / Kiel University}
These three internal German datasets were used for training and testing the TTS quality model in \cite{NORRENBROCK201517}.

Test 1 was carried out to analyze perceptual quality dimensions of TTS systems in \cite{hinterleitner2011perceptual}, which also contains a detailed description of the test. The dataset overall includes 60 files from 15 TTS systems with different speakers (male/female). Ten German sentences were used in this test and the speech material was synthesized by following systems:  Acapela Infovox3, AT\&T Natural Voice, atip Proser, BOSS, Cepstral Voices, Cereproc CereVoice, DRESS, Loquendo, MARY bits, MARY hmm-bits, MARY MBROLA, NextUp Talker, NextUp TextAloud3, Nuance RealSpeak, SVOX, and SyRUB. The speech samples were then rated by 30 test participants in a soundproof booth at the Quality and Usability Lab, TU Berlin. Because the samples were rated on continuous attribute scales, we linearly transformed the naturalness ratings to a range between 1-5. 

Test 2 was used to determine TTS speech quality dimensions through a multidimensional scaling and is described in \cite{hinterleitner2012}. It consists of 57 speech samples from 16/19 female/male German TTS systems (same as Test 1 and additionally ESpeak, Fonix Speech FonixTalk, IVONA, Meridian Orpheus). All speech samples contained the same 5\,sec long German sentence. The samples were rated by 12 listening test participants at the Quality and Usability Lab, TU Berlin. Again, the naturalness ratings were linearly transformed to a range between 1-5.

Test 3 \cite{seget2007} consists of overall 60 samples from 6 German TTS systems (AT\&T Natural Voices, atip Proser, BOSS, Cepstral Voices, DRESS, MARY MBROLA) that spoke 5 different utterances and is described in \cite{NORRENBROCK201517}. All samples were processed with ITU-T G.712 (telephone bandwidth), which leads to a limited audio bandwidth of 300\,Hz - 3400\,Hz. The speech samples were rated by 17 listening test participants at Kiel University.
\subsection{PhySyQX}
PhySyQX \cite{physqx, gupta2017latent} is a database that was created for physiological evaluation of synthesized speech quality and consists of 44 speech samples, of which 36 samples are publicly available\footnote{http://musaelab.ca/resources/} (two of four natural voices are missing). It contains 2 natural voices and 7 commercial TTS systems (Microsoft, Apple, Mary TTS Unit selection \& HMM, vozMe, Google and Samsung) with each four sentences. Besides simultaneous recordings of 62-channel EEG and 60-functional channel fNIRS, the database also contains results of a subjective listening test experiment with 12 test participants, of which we used the naturalness ratings.
\section{Experiments and results}
To train the model, we use Adam optimizer with a learning rate of 0.001. At first, we train the model with the speech quality pre-training database for 24 epochs. After that, we use this pre-trained model to train on the training datasets, without freezing any weights. We ran the training multiple times and then selected the model that obtained the best results in terms of average Pearson's correlation over all validation datasets. Four datasets were held out of training and validation as a test set. Additionally, we repeated the same procedure without speech quality pre-training to analyze the impact of transfer learning on the model accuracy.

 In contrast to most of the previous work published on synthetic speech naturalness prediction, we do not randomly split all available data in training and validation sets. Instead, we keep certain datasets out of training to see how well the model generalizes without being trained on similar data.

The results for all datasets are presented in Table \ref{tab:results}. In the bottom of the table the average and worst case across the validation and test set are shown. The ``Min." column presents the dataset duration in minutes. The abbreviation after the year of the Blizzard and VCC challenge corresponds to the challenge's tasks. The average per-system correlation with transfer learning on the validation set is 0.89 and on the test set 0.77. 

These results are notably higher than the results without transfer learning with correlations of 0.85 and 0.71. In particular, on the datasets ``Blizzard 2012 EH1",  ``Blizzard 2009 MS1/MS2", and ``PhySyQX" the correlation is higher. All of these datasets use speakers that are not contained in the training set. In contrast, for datasets that use different systems but the same speaker (e.g. ``Blizzard 2012 EH2") the correlations are more similar. This indicates that the speech quality pre-training with more than 1,000 speakers improved the speaker-independent naturalness prediction of the proposed model. 

While the per-system results with correlations are overall promising, the average correlation per-stimuli is only 0.65 on the validation set; this shows that the model can only reliably be applied on a system level. However, partly the small number of ratings per-stimuli and, therefore, high confidence intervals of the subjective MOS could influence the results as well. For example, the datasets ``Test 2" and ``PhySyQX" have a higher number of ratings per-stimuli and also a relatively high per-stimuli correlation.

It should be noted that for ``Blizzard 2016 EH", where the per-system correlation is only 0.77, we do not know which speech samples were used in the listening test since no per-stimuli ratings are available. Because of this, we used all available speech samples, some of which may not have appeared in the listening test.

The results on the dataset ``Test 3" of the test set stand out with an extremely low correlation of 0.33. In contrast to the other datasets, this one was post-processed with a telephone bandwidth filter. Therefore, it can be assumed that the model is confused by this particular filter. Although the speech samples of the ES2 task of Blizzard Challenge 2009 were also processed with a telephone bandwidth filter, these are only 338 files in the training set.

\begin{figure}[!ht]

\begin{minipage}[b]{.48\linewidth}
  \centering
  \centerline{\small PhySyQX}\medskip
  \vspace{-0.25cm}
  \centerline{\includegraphics[width=3.6cm]{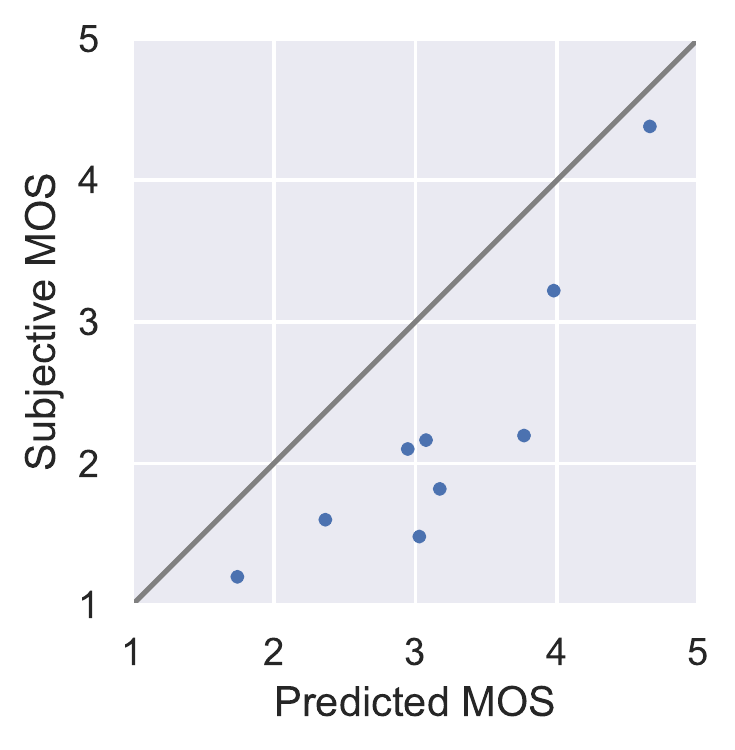}}
\end{minipage}
\hfill
\begin{minipage}[b]{0.48\linewidth}
  \centering
  \centerline{\small Test 2} \medskip
 \vspace{-0.25cm}
  \centerline{\includegraphics[width=3.6cm]{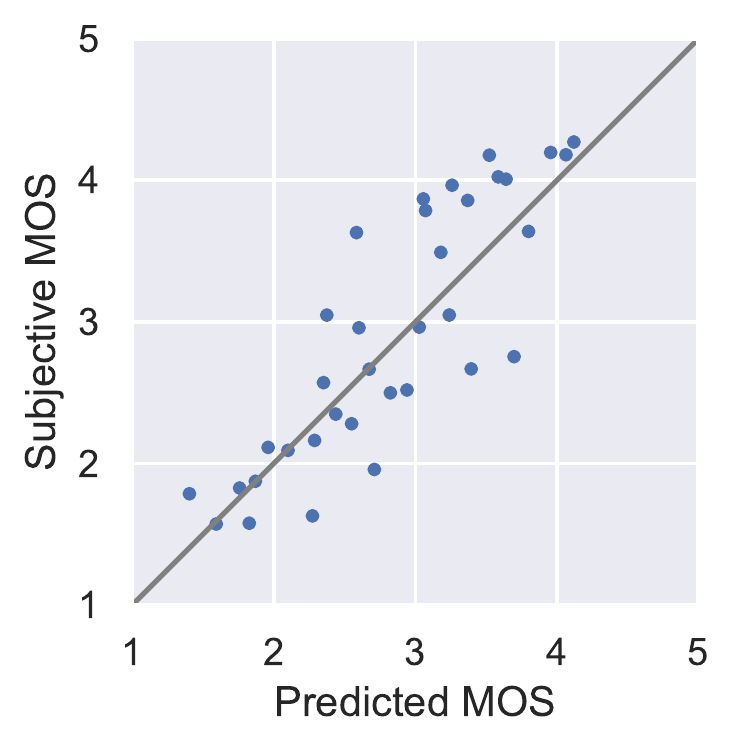}}
\end{minipage}
 \vspace{0.28cm}


\caption{Per-system correlation diagrams of PhySyQX and Test 2 dataset}
\label{fig:reslts}
\end{figure}

Looking at the other datasets of the test set, the results are very promising. ``PhySyQX" is the most independent dataset, as it contains new systems and speakers that were not contained in the training set. Also, the companies that contributed to the dataset produced the samples themselves, some with female others with male voices. Still the model obtains a high correlation of 0.89 on a system level (also see Figure \ref{fig:reslts}). 

MosNet \cite{Lo2019}, was also tested on the VCC 2016 dataset with a correlation of 0.92, while only being trained on the VCC 2018 data set. Our proposed model outperforms these results with a correlation of 0.96. Interestingly, the proposed model without transfer learning achieves a similar correlation of 0.93.
\section{Conclusions}
Our proposed TTS naturalness prediction model achieved promising results on unseen datasets. Particularly when used on a system-level it shows to provide reliable results that could be used during the development of new TTS systems, for example, to compare different configurations. The model works language independently and is made publicly available. However, it showed to be unreliable for speech samples that have been processed with a telephone bandwidth filter, which we will analyze in more detail in future research. In a next step, the pre-training database could be improved further with an even wider variety of different speakers and conditions that are more similar to TTS distortions. Also, the full-reference speech quality model presented in \cite{mittag_b}, which automatically aligns the reference to the degraded signal, could be used to estimate the similarity between original and synthesized/voice-conversed speakers.
\section{Acknowledgements}
The authors would like to thank the organizers of the Blizzard Challenge and the Voice Conversion Challenge for making the results of past challenges available. Also, we would like to thank MuSAE Lab for making the PhySyQX dataset available.

The work on this paper was largely supported by the BMBF, Grant 01IS17052.

The AusTalk corpus was collected as part of the Big ASC project (Burnham et al. 2009; Wagner et al. 2010; Burnham et al. 2011), funded by the Australian Research Council (LE100100211). See: https://bigasc.edu.au/ for details

\bibliographystyle{IEEEtran}

\bibliography{mybib}


\end{document}

%% file: results_table.tex
\begin{table*}[!t]
\caption{Results in terms of Pearson's correlation and Root-Mean-Square Error}
\vspace{-0.3cm}
\centering
\label{tab:results}
\resizebox{0.99\linewidth}{!}{%
\begin{threeparttable}
\begin{tabular}{@{}ll|cccccc|cccc|cccc@{}}
 &  &\multicolumn{4}{c}{}  & &  & \multicolumn{4}{c}{Without Transfer Learning} & \multicolumn{4}{c}{With Transfer Learning} \\
 &  & \multicolumn{2}{c}{\textit{}} & & \multicolumn{2}{c}{\textit{}} &  & \multicolumn{2}{c}{Per Stimuli} & \multicolumn{2}{c}{Per System} & \multicolumn{2}{c}{Per Stimuli} & \multicolumn{2}{c}{Per System} \\
\textit{} & Dataset & Speaker & Lang. & Min. & Files/Sys. & \# Sys. & \# Files & \textit{r} & RMSE & \textit{r} & RMSE & \textit{r} & RMSE & \textit{r} & RMSE \\ \toprule \toprule
Training & Blizzard 2008 A & Roger & en & 42 & 42.0 & 21 & 882 & 0.95 & 0.28 & 0.99 & 0.07 & 0.95 & 0.28 & 1.00 & 0.07 \\
 & Blizzard 2008 B & Roger &  en & 36 & 38.0 & 19 & 722 & 0.95 & 0.28 & 0.99 & 0.09 & 0.95 & 0.28 & 1.00 & 0.08 \\
 & Blizzard 2009 EH1 & Roger &  en & 89 & 37.8 & 18 & 681 & 0.95 & 0.30 & 0.99 & 0.11 & 0.95 & 0.29 & 0.99 & 0.11 \\
 & Blizzard 2009 EH2 & Roger &  en & 23 & 19.0 & 19 & 361 & 0.95 & 0.28 & 0.99 & 0.13 & 0.95 & 0.30 & 0.99 & 0.13 \\
 & Blizzard 2009 ES1/ES2 & Roger &  en & 57 & 22.2 & 21 & 466 & 0.93 & 0.35 & 0.99 & 0.13 & 0.90 & 0.43 & 0.97 & 0.22 \\
 & Blizzard 2010 EH1 & RJS &  en & 31 & 36.0 & 18 & 648 & 0.96 & 0.30 & 0.99 & 0.16 & 0.95 & 0.33 & 1.00 & 0.16 \\
 & Blizzard 2010 ES1/ES3 & RJS & en & 13 & 16.0 & 16 & 256 & 0.96 & 0.32 & 0.99 & 0.19 & 0.96 & 0.31 & 1.00 & 0.12 \\
 & Blizzard 2010 MH1 & CAS & zh & 23 & 24.0 & 12 & 288 & 0.96 & 0.30 & 0.99 & 0.15 & 0.94 & 0.33 & 0.99 & 0.10 \\
 & Blizzard 2010 MH2 & CAS & zh & 25 & 33.0 & 11 & 363 & 0.96 & 0.31 & 0.99 & 0.16 & 0.95 & 0.32 & 0.99 & 0.11 \\
 & Blizzard 2010 MS1 & CAS & zh & 5 & 12.0 & 6 & 72 & 0.96 & 0.31 & 0.99 & 0.15 & 0.96 & 0.29 & 0.99 & 0.19 \\
 & Blizzard 2011 EH1 & Nancy & en & 45 & 39.0 & 13 & 507 & 0.95 & 0.28 & 0.99 & 0.10 & 0.94 & 0.29 & 0.99 & 0.09 \\
 & Blizzard 2013 EH1 & C. Byers & en & 52 & 51.0 & 10 & 510 & 0.96 & 0.30 & 0.99 & 0.13 & 0.96 & 0.30 & 1.00 & 0.08 \\
 & Blizzard 2013 EH2 & C. Byers &  en & 73 & 52.4 & 15 & 786 & 0.95 & 0.32 & 0.99 & 0.14 & 0.94 & 0.31 & 0.99 & 0.09 \\
 & Blizzard 2014 IH*† &  &  as, gu, hi, rj, ta, te & 981 & 68.6 & 131 & 8990 & N/A & N/A & 0.99 & 0.11 & N/A & N/A & 0.99 & 0.13 \\
 & Blizzard 2015 IH*† &  & bn, ml, hi, mr, ta, te & 499 & 50.0 & 104 & 5200 & N/A & N/A & 0.96 & 0.12 & N/A & N/A & 0.94 & 0.15 \\
 & Blizzard 2019 MH & Z. Luo  & zh & 186 & 52.0 & 26 & 1352 & 0.98 & 0.18 & 1.00 & 0.11 & 0.99 & 0.17 & 1.00 & 0.09 \\
 & Test 1 &  &  de & 10 & 2.0 & 30 & 60 & 0.93 & 0.27 & 0.96 & 0.21 & 0.89 & 0.33 & 0.92 & 0.28 \\
 & VCC 2018 HUB Train &  &  en & 55 & 38.5 & 26 & 1000 & 0.59 & 0.77 & 0.98 & 0.14 & 0.64 & 0.72 & 0.98 & 0.15 \\
 & VCC 2018 SPO Train &  &  en & 56 & 71.4 & 14 & 1000 & 0.66 & 0.76 & 0.97 & 0.18 & 0.70 & 0.71 & 0.99 & 0.12 \\ \midrule
Validation & Blizzard 2008 C &  CAS & zh & 25 & 26.0 & 13 & 338 & 0.62 & 0.71 & 0.87 & 0.37 & 0.75 & 0.57 & 0.90 & 0.34 \\
 & Blizzard 2009 MH & iFLYTEK & zh & 38 & 24.0 & 12 & 288 & 0.49 & 1.05 & 0.77 & 0.85 & 0.57 & 0.89 & 0.79 & 0.72 \\
 & Blizzard 2009 MS1/MS2 & iFLYTEK & zh & 22 & 16.3 & 16 & 260 & 0.51 & 0.94 & 0.73 & 0.73 & 0.64 & 0.76 & 0.84 & 0.57 \\
 & Blizzard 2010 EH2 & Roger & en & 41 & 36.0 & 18 & 648 & 0.67 & 0.71 & 0.91 & 0.36 & 0.72 & 0.66 & 0.94 & 0.26 \\
 & Blizzard 2012 EH1 & J. Greenman & en & 33 & 40.2 & 11 & 442 & 0.64 & 1.10 & 0.77 & 0.93 & 0.63 & 0.81 & 0.88 & 0.60 \\
 & Blizzard 2016 EH* & L. Sims & en & 284 & 68.0 & 17 & 1156 & N/A & N/A & 0.76 & 0.79 & N/A & N/A & 0.77 & 0.60 \\
 & VCC 2018 HUB Val & & en & 53 & 38.5 & 26 & 1000 & 0.59 & 0.76 & 0.96 & 0.18 & 0.61 & 0.73 & 0.99 & 0.14 \\
 & VCC 2018 SPO Val & & en & 56 & 71.4 & 14 & 1000 & 0.66 & 0.74 & 0.99 & 0.12 & 0.68 & 0.71 & 0.99 & 0.15 \\ \midrule
Test & Test 2 & & de & 5 & 1.6 & 35 & 57 & 0.61 & 0.75 & 0.71 & 0.66 & 0.80 & 0.52 & 0.85 & 0.47 \\
 & Test 3 & & de & 13 & 5.0 & 12 & 60 & 0.34 & 0.92 & 0.38 & 0.86 & 0.33 & 0.96 & 0.38 & 0.92 \\
 & VCC 2016* & & en & 1428 & 1301.4 & 20 & 26028 & N/A & N/A & 0.93 & 0.51 & N/A & N/A & 0.96 & 0.37 \\
 & PhySyQX & & en & 12 & 4.0 & 9 & 36 & 0.72 & 1.11 & 0.83 & 1.00 & 0.82 & 1.10 & 0.89 & 1.04 \\ \bottomrule \bottomrule
Average & Validation & &  &  &  &  &  & 0.60 & 0.86 & 0.85 & 0.54 & 0.65 & 0.73 & 0.89 & 0.42 \\
 & Test & &  &  &  &  &  & 0.58 & 0.89 & 0.71 & 0.76 & 0.68 & 0.79 & 0.77 & 0.70 \\
Worst Case & Validation & &  &  &  &  &  & 0.49 & 1.10 & 0.73 & 0.93 & 0.57 & 0.89 & 0.77 & 0.72 \\
 & Test & &  &  &  &  &  & 0.34 & 1.11 & 0.38 & 1.00 & 0.33 & 1.10 & 0.38 & 1.04
\end{tabular}
\begin{tablenotes}\footnotesize
\item[*] No Per-Stimuli ratings available
\item[†] Consisted of 6 languages with each 3 tasks, summarized to one dataset
\end{tablenotes}
\end{threeparttable}
}
\end{table*}